\documentclass[10pt]{wlscirep}
\usepackage{hyperref}
\usepackage{graphicx}
\usepackage{amsmath,amsfonts,amsthm,amssymb}
\usepackage{pdflscape}
\usepackage[separate-uncertainty,table-align-uncertainty,multi-part-units=repeat]{siunitx}
\newcolumntype{C}[1]{>{\centering\let\newline\\\arraybackslash\hspace{0pt}}p{#1}}
\usepackage[utf8]{inputenc}
\usepackage[T1]{fontenc}
\usepackage{textgreek}

\title{Assessing Left Main Bifurcation Anatomy and Haemodynamics: A Potential Surrogate for Disease Risk in Suspected Coronary Artery Disease Without Stenosis?}

\author[1,*]{R. Gharleghi}
\author[1]{M. Zhang}
\author[1]{C. Shen}
\author[2]{M. Webster}
\author[2]{C. Ellis}
\author[1]{S. Beier}
\affil[1]{School of Mechanical and Manufacturing Engineering, UNSW, Sydney, Australia}
\affil[2]{Auckland City Hospital, Auckland, New Zealand}

\affil[*]{Corresponding Author: r.gharleghi@unsw.edu.au}

\begin{abstract}
Coronary anatomy governs local haemodynamics associated with atherosclerotic development, progression and ultimately adverse clinical outcomes. However, lack of large sample size studies and methods to link adverse haemodynamics to anatomical information has hindered meaningful insights to date. 

The Left Main coronary bifurcations of 127 patients with suspected coronary artery disease in the absence of significant stenosis were segmented from CTCA images before computing the local haemodynamics. We correlated 11 coronary anatomical characteristics with the normalised lumen area exposed to adverse haemodynamics linked with atherosclerotic processes. These include mean curvatures and diameters of branches, bifurcation and inflow angles, and Finet's ratio as the anatomical parameters, and low Time-Averaged Endothelial Shear Stress (lowTAESS\textless \SI{0.5}{Pa}), high Oscillatory Shear Index (highOSI\textgreater 0.1), and high Relative Residence Time (highRRT\textgreater \SI{4.17}{\per\pascal}) for the haemodynamic consideration. We separately tested if the geometric measures and haemodynamics indicators differed between subgroups (sex, smokers, and those with hypertension). We then use a step-down multiple linear regression model to find the best model for predicting lowTAESS, highOSI and highRRT.

Finet's Ratio (FR) significantly correlated to lowTAESS (p\textless 0.001). Vessel diameters and curvature correlated to highOSI (both p\textless 0.001). Finet's ratio, vessel diameters and daughter branch curvature independently correlated to RRT (all p\textless 0.01).

Our results indicate that specific anatomical vessel characteristics may be used as a surrogate of adverse haemodynamic environment associated with clinically adverse mechanisms of disease. This is especially powerful with the latest computing resources and may unlock clinical integration via standard imaging modalities as biomarkers without further computationally expensive simulations. 
\end{abstract}
\begin{document}

\flushbottom
\maketitle
\thispagestyle{empty}
\section*{Introduction}
Coronary artery disease, a leading cause of death worldwide, is closely associated with changes in vessel physiology such as atherosclerotic plaques narrowing the coronary arteries, with subsequent restriction of the coronary blood supply. Local blood flow dynamics, particularly Endothelial Shear Stress (ESS), have been shown to regulate signalling pathways driving such morphological changes, e.g. endothelial cell growth leading to neointimal thickening \cite{malek1999hemodynamic,davies1995flow}. The anatomy of coronary arteries, represented by parameters such as vessel calibre, angulation, and curvature, directly influences the blood flow and, thus, the distribution and magnitude of ESS. This forms the ``anatomy of risk'' hypothesis, whereby the blood flow-induced shear forces, called haemodynamics, affect disease mechanisms - thus being directly associated with vessel shape characteristics. Bifurcating coronary segments are more likely to be affected by disease \cite{cecchi2011role}, which can be attributed to their complex blood flow dynamics with both adversely high and low shear regions due to recirculation, stagnating, and rapid blood flow regions initiated due to the flow split from the main vessel into the smaller arterial segments \cite{cecchi2011role}.  Understanding these mechanisms may prove vital for risk stratification and ultimately preventative measures,  especially in the Left Main (LM), which supplies the majority of cardiac tissue \cite{loop1979atherosclerosis}.  

Thus, vascular anatomy has repeatedly been suggested as a surrogate for adverse haemodynamic computations in individuals \cite{zhu2009cataloguing, morbiducci2016atherosclerosis},  since these could be identified in medical images directly.  However, until recently, limited spatial resolution and a 2D rather than 3D standard imaging have hindered the effective capture of complex coronary anatomical characteristics. This is particularly apparent compared to related, more advanced fields such as carotid research, since carotid arteries are comparatively easier to image as they are larger, stagnant, and close to the body surface \cite{beier2016dynamically, taylor2004experimental}. Consequently, the detection of shape biomarkers for the carotid arteries is further developed and was demonstrated to directly correlate to shape features such as the angle of origin \cite{sitzer2003internal}, and vessel curvature \cite{zhang2012flow}, showcasing the distinct relationship between vessel anatomy and adverse flow \cite{lee2008geometry}. In fact, carotid anatomical characteristics have been used as actual age marker \cite{fedintsev2017markers}, and disease risk factor \cite{o1996thickening,liebeskind2015carotid}, which has been not accomplished for the coronaries to date but demonstrates the potential of such research efforts.  

For the coronaries, anatomical characteristics have interdependent effects \cite{beier2016impact, shen2021secondary}, however, studies to date are limited to the consideration of a maximum of three or fewer anatomical parameters at once \cite{rabbi2020computational}. The adverse effect of tortuosity was reported \textit{in vivo} repeatedly \cite{li2011clinical, tuncay2018non}, yet could not be confirmed with computational efforts \cite{li2017impact} – likely due to the lack of consistent 3D measurement methods and mathematical definition \cite{kashyap2022accuracy}, as well as a lack of the consideration of personalised geometries \cite{xie2014computation, song2021numerical}. Vessel diameter has not been extensively considered before, although its critical haemodynamics effect in combination with other shape characteristics has been demonstrated \cite{muller2012pressure, shen2022helical}. Moreover, the small  number of patient-specific geometries tested (n\textless 25) is another gap in the literature to date \cite{rabbi2020computational,pinho2019impact,shen2021secondary}, especially considering the large anatomical variation found within a large population \cite{medrano_gracia2016study}. No other large-scale, patient-specific effort has attempted to link all prominent coronary arterial shape characteristics and adverse haemodynamics before this work as presented here.

Therefore, our aim is to unravel the relationship between coronary anatomy and adverse haemodynamics in a large-scale, patient-specific population (n=127) using all potential anatomical bifurcation factors identified  (n=11) to explore the feasibility of using coronary shape characteristics as biomarkers of adverse blood flow in the LM. 

\section*{Methods}
We collected 127 CTCA images of patients (males 42 / 33\%, females 85/ 67\%, age median 57, age range 38-81, Table \ref{tab:demos}) with suspected coronary artery disease for virtual reconstruction of the LM coronary bifurcation and subsequent computational haemodynamic analysis using transient Computational Fluid Dynamics (CFD) modelling. 

\subsection*{Patient Inclusion and Medical Imaging Protocol}
This study was approved by the local institutional ethics committees of the University of Auckland (Ref. 022961) and the University of New South Wales (Ref. HC190145), with informed and written consent received from all participants. A retrospective review of patients referred to Intra Care (Auckland, NZ) was conducted to include those with no stenosis or calcification within the entire epicardial coronary arteries. We further excluded patients if they had intermediate arteries or had CTCA images of insufficient quality for segmentation or 3D vascular reconstruction due to respiratory or motion artefacts. CTCA images were obtained using a multi-detector CT scanner (GE Lightspeed 64 multi-slice scanner, USA) following a retrospective ECG gating protocol. The contrast-enhanced images were obtained through intravenous administration of a non-ionic medium (Omnipaque 350, GE Healthcare). Where necessary, a beta blocker was used to achieve a resting heart rate of 60 beats per minute to reduce motion artefact.

\begin{table}[h]
\centering
\caption{Patient demographics, coronary anatomical and haemodynamic characteristics.\\
Values are represented as mean \textpm  standard deviation unless otherwise specified.
Adverse haemodynamics refer to the absolute value and normalised percentage luminal area exposed to low time-averaged Endothelical Shear Stress (lowTAESS\textless0.5 Pa), high Oscillatory Shear Index (highOSI\textgreater0.1), and high Relative Residence Time (RRT, \textgreater \SI{4.17}{\per\pascal}) respectively.}
\label{tab:demos}
\begin{tabular}{llccc}
\rowcolor{gray!75}
\multicolumn{2}{l}{Characteristics} & Count \textpm SD       & Proportion & Coefficient of variation \\\hline
\rowcolor{gray!25}
\multicolumn{4}{l}{Patient Demographics }                     &                          \\\hline
Sex                  & Female       & 85            & 66.9\%       &                          \\
                     & Male         & 42            & 33.1\%       &                          \\
Age (years)          & \multicolumn{1}{l}{38-81 (median 57)}            &    127        &                          \\
Ethnicity            & Caucasian    & 111           & 87.4\%    &                          \\
                     & Maori        & 1             & 0.8\%     &                          \\
                     & Asian        & 2             & 1.6\%     &                          \\
                     & Indian       & 5             & 3.9\%     &                          \\
                     & Other        & 8             & 6.3\%     &                          \\
Weight (kg)          &              & 77.89 \textpm 14.39 &            &                          \\
Height (m)           &              & 1.68 \textpm 0.10   &            &                          \\
BMI (kg/m2)&              & 27.39 \textpm 4.35  &            &                          \\
Smoker               &              & 41            & 32\%       &                          \\
Hypertension         &              & 36            & 28\%       &                          \\
Diabetes             &              & 3             & 2\%        &                          \\\hline
\rowcolor{gray!25}
\multicolumn{4}{l}{Coronary Anatomy}                        &                          \\\hline
Diameter [mm]       & LM           & 3.56 \textpm 0.66   &            & 18.7\%                   \\
                     & LAD          & 3.21 \textpm 0.57   &            & 17.9\%                  \\
                     & LCx          & 3.08 \textpm 0.63   &            & 20.6\%                  \\
Curvature [\si{\per\mm}]~~& LM           & 0.44 \textpm 0.11   &       & 25.7\%               \\
                     & LAD          & 0.49 \textpm 0.10   &            & 20.7\%                  \\
                     & LCx          & 0.51 \textpm 0.12   &            & 23.7\%                  \\
Angle A (°)          &              & 124 \textpm 20      &            & 16.2\%                  \\
Angle B (°)          &              & 79 \textpm 19       &            & 23.7\%                  \\
Inflow Angle (°)     &              & 17 \textpm 15       &            & 88.7\%                  \\
Finet's Ratio        &              & 0.57 \textpm 0.03   &            & 5.8\%                   \\\hline
\rowcolor{gray!25}
\multicolumn{4}{l}{Adverse Haemodynamics}                  &                          \\\hline
lowTAESS@0.5 (\%)       &              & 25.62 \textpm 9.36  &            &                       \\
highOSI@0.1 (\%)        &              & 3.08 \textpm 2.36   &            &                       \\
highRRT@4.17 (\%)        &              & 4.84 \textpm 3.06   &            &           \\\bottomrule
\end{tabular}
\end{table}

\subsection*{Coronary Segmentation and Reconstruction}
The image sets were semi-automatically segmented as previously described \cite{medrano_gracia2014construction}. Briefly, an experienced analyst used standard software (Osirix CMIV CTA version 4.1.2 32-bit and MiaLight version 1.0) for segmentation, generating the surface mesh and vessel centerlines semi-automatically taking approximately 20 minutes per case. The resulting geometries were then imported into the open-source Vascular Modelling Tool Kit (VMTK) \cite{piccinelli2009framework} for automatic smoothing (Taubin \cite{taubincurve} filter with a passband of 0.03 cut-off and 30 iterations), and adding of flow extensions (4x radius perpendicular to the median direction of the last 5mm) before meshing using patch-conforming, unstructured tetrahedral elements with five boundary layers. 

\subsection*{Quantification of Coronary Geometric Indices}
\begin{figure}[h]
\centering
\includegraphics[width=0.8\linewidth]{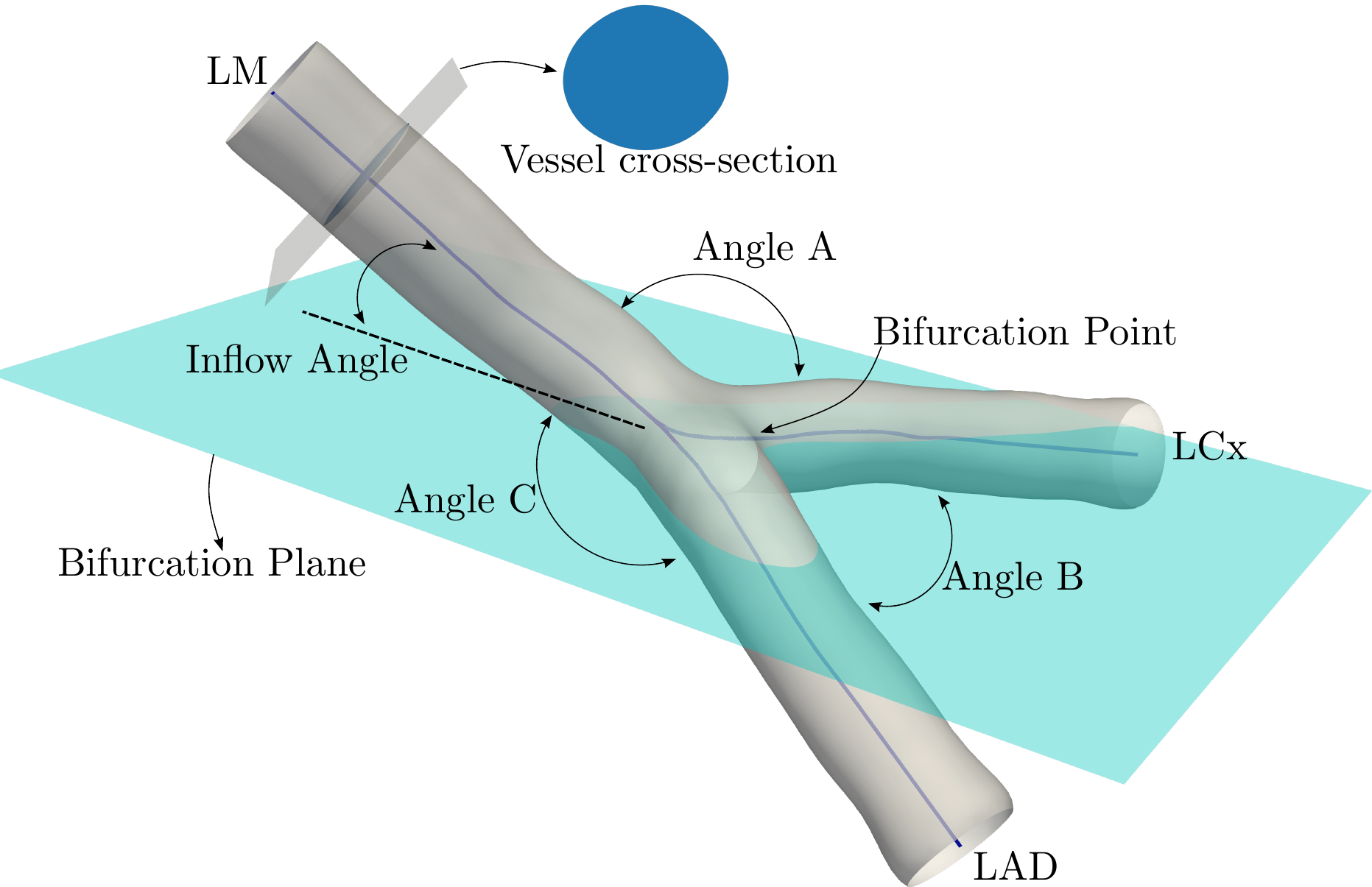}
\caption{Schematic of coronary anatomical metrics, including angles A, B, and C, inflow angle, Left Main (LM), Left Circumflex (LCx), Left Anterior Descending (LAD) and the bifurcation point.}
\label{fig:anatomy}
\end{figure}

The LM bifurcation section within 10 mm of the bifurcation point was extracted and the shape characteristics were automatically quantified using an in-house software \cite{beier2016impact} according to the European Bifurcation Club's naming convention \cite{lassen2014percutaneous}: (i) Angle A: the angle between LM and LAD, (ii) Angle B: the angle between LAD and LCx, (iii) Inflow Angle: the angle between LM and the plane formed by the LAD and LCx as a measure of how acute or obtuse the LM branches from the aortic arch, (iv) tortuosity of each vessel, defined as the mean curvature of the centreline, which was found to be superior to the tortuosity index based on prior work \cite{kashyap2022accuracy}, (v) median diameter for each vessel, and (vi) Finet's ratio, defined as (LM diameter)/(LAD diameter + LCx diameter). Figure \ref{fig:anatomy} summarises the relevant geometric features.

\subsection*{Computational Fluid Dynamics}
The vascular walls were modelled as rigid bodies since a previous computational study showed little haemodynamic difference between rigid-wall and compliant fluid-structure interaction simulations in cycle averaged metrics \cite{eslami2020effect}. A tetrahedral grid with five prism layers was generated for the blood flow domain. After a sensitivity analysis, the mesh cardinality was approximately 2x10$^6$ with a time-step of 0.002 seconds. The Navier-Stokes equations governing the laminar fluid flow were resolved using CFX (version 19.3, ANSYS, USA). Flow profile in coronary arteries is blunted due to the pulsatile nature of the flow \cite{tangasme}, hence, a uniform velocity profile was prescribed at the inlet. A zero reference static pressure condition was defined at both outlets in line with experimental observations \cite{beier2016dynamically}, and deemed suitable for non-diseased cases. Four consecutive cardiac cycles were simulated, with the results taken from the fourth cycle to minimise the transient start-up effects. A residual convergence criterion of 10$^{-4}$ was used. A standard non-slip boundary condition at the vessel walls and shear-thinning Carreau-Yasuda model were applied \cite{razavi2011numerical}. Computations were performed on the UNSW Katana cluster, taking approximately 25 CPU-hours per bifurcation. 

\subsection*{Quantification of Adverse Haemodynamics}
We analysed TAESS, OSI, and RRT. LowTAESS \cite{he1996pulsatile} quantifies exposure to insufficient shear stress, highOSI \cite{hoi2011correlation} areas with changes in the direction of ESS, and highRRT \cite{hoi2011correlation} attempts to combine both metrics, thereby capturing both low and oscillating shear stress. These are calculated as follows:
\begin{align}
    \text{TAESS} &= \frac 1 T \int_0^T{|\mathbf{\tau}_w}|dt\\
    \text{OSI} &= \frac 1 2 \left( 1- \frac{|\int_0^T{\mathbf{\tau}_w}dt|}{\int_0^T{|\mathbf{\tau}_w}|dt}\right)\\
    \text{RRT} &= \frac{1}{(1-2\cdot\text{OSI})\cdot\text{TAESS}}
\end{align}

Where $\mathbf{\tau}_w$ is the flow-induced shear stress vector at the arterial vessel wall, and\textit{ T} denotes the cardiac cycle period. Adverse thresholds as per literature include lowTAESS\textless \SI{0.5}{Pa} \cite{rabbi2020computational,xie2014computation}, and highOSI\textgreater 0.1 \cite{xie2014computation}, which have both been associated with neointimal thickening and plaque formation. While the threshold for RRT can be calculated from the TAESS and OSI thresholds (resulting in \textgreater \SI{2.5}{\per\pascal} in this case), we chose to use a threshold of \textgreater \SI{4.17}{\per\pascal} as commonly used in literature \cite{rabbi2020computational}. We note that these thresholds are not universal. However, we investigated whether different choices of thresholds (lowTAESS\textless \SI{0.4}{Pa}, lowTAESS\textless \SI{0.6}{Pa}, highOSI\textgreater 0.2) affected the overall findings and found negligible differences (see Table \ref{tab:TAESSThreshold} and \ref{tab:OSIThresholds} in the Supplementary Material). Other haemodynamic parameters such as ESS Gradient (ESSG), or instantaneous ESS have also been used in the past, yet their impact is more ambiguous \cite{soulis2014wall}. Moreover,  the implications of instantaneous haemodynamics throughout the cardiac cycle have never been established. We normalised and computed the percentage of area exposed to the LM arterial wall across all cases.

\subsection*{Statistical Analyses}
Statistical analyses were performed using the python statsmodels \cite{seabold} package. Continuous and categorical variables were expressed as mean and Standard Deviation (SD), and counts and percentages, respectively. Independent two-sided Welch's t-tests were used to examine the differences in each anatomical or haemodynamic metric between sexes and between patients with and without hypertension or smoking history. This test is robust to violations of homoscedasticity (where the subgroups have unequal variance) \cite{ruxton2006unequal} and violations of normality assumption \cite{delacre2017why}. Analysis for diabetes was not conducted due to the limited sample size (n=3). 

Correlations between each of the shape metrics and the normalised vessel area exposed to lowTAESS, highOSI, and highRRT were examined using standardised regression coefficients, i.e. the variables were standardised to have unit variance such that the regression coefficient measures how many standard deviations the output will change in response to a one-standard-deviation change in the predictor. The obtained p-values were adjusted using Holm-Bonferroni correction \cite{abdi2010holm} to prevent inflated false positive rates due to multiple comparisons. A p-value \textless 0.05 was considered statistically significant throughout this work. 

To analyse the predictive performance of the studied metrics, a backward multiple regression procedure \cite{yamashita2007stepwise} was used to find a set of predictors with the lowest Akaike Information Criterion (AIC) \cite{hu2007akaike}. This balances prediction performance and model simplicity, avoiding overly complex models.

\section*{Results}
\subsection*{Variations in Coronary Anatomy and Haemodynamics}
Coronary anatomy varied widely across the 127 patients (Table \ref{tab:demos}, partly previously reported elsewhere \cite{medrano_gracia2016study}, but we include curvature,  exclude intermediate arteries and apply a Bonferroni correction here), with the inflow angle showing the largest variation from -40 to 45° with a coefficient of variation of 88.7\%. Other parameters also varied markedly, including the curvature of the LM (25.7\%) and the LCx (23.7\%). Notably, Finet's ratio showed a very small variation of only 5.8\% across all patients (Table \ref{tab:demos}).

After adjusting for multiple comparison testing, there were no statistically significant differences between patients with and without hypertension, and smokers and non-smokers. 
Females had significantly smaller diameters in all LM branches compared to males (mean difference, LM 0.69 mm, 95\% CI: 0.48 to 0.898, p\textless 0.001; LAD: 0.60 mm, 95\% CI: 0.41 to 0.79, p\textless 0.001; LCx: 0.63 mm, 95\% CI: 0.42 to 0.84, p\textless 0.001). The difference in vessel diameter persisted after adjusting for Body Mass Index and Body Surface Area. Females showed higher curvature in all vessels with a difference of LM: \SI{0.062}{\per\mm} (95\% CI: 0.025 to 0.099, p\textless 0.001); LAD: \SI{0.063}{\per\mm} (95\% CI: 0.031 to 0.096, p\textless 0.001); and LCx: \SI{0.072}{\per\mm} (95\% CI: 0.034 to 0.84, p\textless 0.001). While the Inflow Angle seemed larger in females, this difference was not significant after adjusting for multiple comparison testing. The statistical results for hypertension, smoking and sex are included in the Supplementary Material (Tables \ref{tab:sexsupp}-\ref{tab:smokingsupp}).

Strong variations were found for all haemodynamic metrics: lowTAESS@0.5 (25.62 ± 9.36 \%), highOSI@0.1 (3.08 ± 2.36 \%), and highRRT@4.17 (4.84 ± 3.06 \%). Subgroup analyses did not  show statistically significant effect of hypertension or smoking history on adverse haemodynamic metrics. While there was no significant difference in lowTAESS between males and females, both highOSI@0.1 (4.6\% vs. 2.35\%, p\textless 0.001) and highRRT@4.17 (6.1\% vs 4.2\%, p=0.01) were higher in males.

\subsection*{Correlations between Anatomy and Coronary Haemodynamics}
\begin{figure}[h]
\centering
\includegraphics[width=0.6\linewidth]{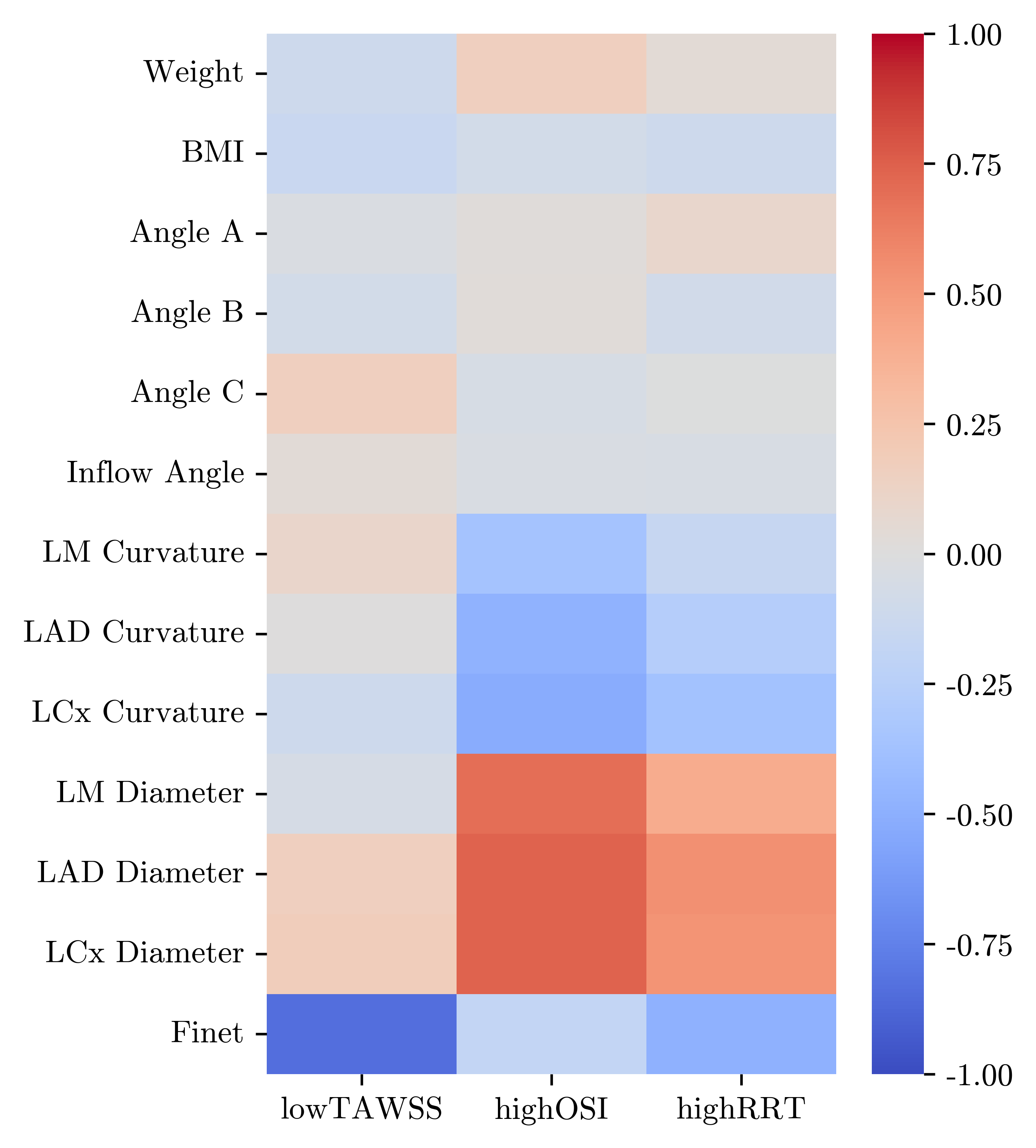}
\caption{Correlogram of the coronary anatomical metrics and the percentage of arterial wall area exposed to adversely low Time-Averaged Endothelial Shear Stress (lowTAESS\textless0.5Pa), high Oscillatory Shear Index (highOSI \textgreater0.1) and Relative Residence Time (highRRT\textgreater \SI{4.17}{\per\pascal}) with dark red (1) being highly positively correlated, dark blue (-1) being highly inversely correlated and grey (0) no correlation.}
\label{fig:corr}
\end{figure}

For lowTAESS only the Finet's ratio correlation was statistically significant (\textbeta = -0.84, p \textless 0.001). For highOSI, both diameters (LM: \textbeta =0.69, p\textless 0.001; LAD: \textbeta =0.74, p\textless 0.001; and LCx: \textbeta =0.74, p\textless 0.001) and curvature (LM: \textbeta =-0.367, p\textless 0.001; LAD: \textbeta =-0.484, p\textless 0.001; and LCx: \textbeta =-0.519, p\textless 0.001) had a statistically significant effect. The Finet's ratio was not statistically significantly correlated to highOSI after adjusting for multiple comparisons. For highRRT, also diameters (LM: \textbeta =0.403, p\textless 0.001; LAD: \textbeta =0.542, p\textless 0.001; and LCx: \textbeta =0.525, p\textless 0.001), daughter branch curvatures (LAD: \textbeta =-0.269, p=0.018; and LCx: \textbeta =-0.368, p\textless 0.001), and the Finet's ratio (\textbeta =-0.488, p \textless 0.001) showed a statistically significant correlation. Correlations are shown in Fig. \ref{fig:corr} and Table \ref{tab:haemoresults}.

{\setlength{\tabcolsep}{2pt}
\begin{table}[ht]
\centering
\caption{Independent tests of effects of patient weight, BMI and geometric features on adverse haemodynamic metrics with * indicating the significance of p\textless0.05.}
\label{tab:haemoresults}
\rowcolors{2}{gray!25}{white}
\begin{tabular}{l|ccC{1.65cm}|ccC{1.65cm}|ccC{1.65cm}}
                & \multicolumn{3}{c}{\textbf{lowTAESS@0.5}} & \multicolumn{3}{c}{\textbf{highOSI@0.1}}         & \multicolumn{3}{c}{\textbf{highRRT@4.17}}       \\
\textbf{Feature}         & \textbf{Coefficient} & \textbf{p-value      }    & \textbf{Adjusted p-value}  & \textbf{Coefficient} & \textbf{p-value} & \textbf{Adjusted p-value} & \textbf{Coefficient} & \textbf{p-value}          & \textbf{Adjusted p-value} \\\midrule
Weight          & -0.111       & 0.215      & 1          & 0.164  & 0.066            & 0.397             & 0.044  & 0.626            & 1                 \\
BMI             & -0.147       & 0.099      & 0.890      & -0.082 & 0.362            & 1                 & -0.110 & 0.218            & 1                 \\
Angle A         & -0.028       & 0.753      & 1          & 0.020  & 0.826            & 1                 & 0.090  & 0.315            & 1                 \\
Angle B         & -0.078       & 0.381      & 1          & 0.027  & 0.766            & 1                 & -0.087 & 0.329            & 1                 \\
Angle C         & 0.161        & 0.071      & 0.780      & -0.048 & 0.593            & 1                 & -0.007 & 0.941            & 1                 \\
Inflow   Angle  & 0.034        & 0.702      & 1          & -0.038 & 0.673            & 1                 & -0.046 & 0.606            & 1                 \\
LM   Curvature  & 0.094        & 0.294      & 1          & -0.367 & \textless{}0.001 & \textless{}0.001* & -0.157 & 0.079            & 0.552             \\
LAD   Curvature & 0.006        & 0.948      & 1          & -0.484 & \textless{}0.001 & \textless{}0.001* & -0.269 & 0.002            & 0.018*            \\
LCX   Curvature & -0.112       & 0.208      & 1          & -0.519 & \textless{}0.001 & \textless{}0.001* & -0.368 & \textless{}0.001 & \textless{}0.001* \\
LM   Diameter   & -0.061       & 0.494      & 1          & 0.691  & \textless{}0.001 & \textless{}0.001* & 0.403  & \textless{}0.001 & \textless{}0.001* \\
LAD   Diameter  & 0.160        & 0.072      & 0.780      & 0.738  & \textless{}0.001 & \textless{}0.001* & 0.542  & \textless{}0.001 & \textless{}0.001* \\
LCX   Diameter  & 0.172        & 0.053      & 0.634      & 0.739  & \textless{}0.001 & \textless{}0.001* & 0.525  & \textless{}0.001 & \textless{}0.001* \\
Finet's   Ratio & -0.842      & \textless{}0.001 & \textless{}0.001* & -0.187      & 0.035   & 0.247            & -0.488      & \textless{}0.001 & \textless{}0.001*
\end{tabular}

\end{table}
}

\subsection*{Multiple regression model}
The diameters for LAD and LCx were pre-emptively removed from the set of predictors to avoid high multi-collinearity in the multiple regression model. At each iteration, one predictor was removed that would result in the set of predictors with the lowest AIC, until AIC could not be decreased by removing predictors. The final retained predictors for each haemodynamic metric are shown in Table \ref{tab:mlr}, whereby the regression coefficient signifies the amount by which a change in the respective haemodynamic metric must be multiplied to give the corresponding average change in the retained feature, or the amount of which the retained feature changes for a unit increase in the haemodynamic metric. This means a larger value indicates a greater effect, with a positive/negative correlation according to the coefficient. 

Our results indicate that a large percentage of the local haemodynamic can be explained with coronary shape characteristics, including 79\% of the variance in lowTAESS, and 63\% and 58\% of the variance for highOSI and high RRT, respectively (Table \ref{tab:mlr}).

\begin{table}[h]
\centering
\caption{Multiple linear regression between coronary anatomy and haemodynamics.}
\label{tab:mlr}
\rowcolors{2}{gray!25}{white}
\begin{tabular}{lc|lc|lc}
\multicolumn{2}{c}{\textbf{lowTAESS (R$^2$ = 0.791)}} & \multicolumn{2}{c}{\textbf{highOSI (R$^2$ = 0.633)}} & \multicolumn{2}{c}{\textbf{highRRT (R$^2$ = 0.585)}} \\
\textbf{Retained Features}  & \textbf{Coefficient}  & \textbf{Retained Features}  & \textbf{Coefficient} & \textbf{Retained Features}   & \textbf{Coefficient}  \\\midrule
Angle A         & 0.259  & BMI           & -0.15  & BMI           & -0.133 \\
Angle C         & 0.166  & Angle A       & 0.097  & Angle A       & 0.243  \\
LM   Curvature  & -0.15  & LAD Curvature & 0.16   & LAD Curvature & 0.155  \\
LCx   Curvature & 0.129  & LM Diameter   & 0.924  & LM Diameter   & 0.692  \\
LM   Diameter   & 0.173  & Finet's Ratio & -0.389 & Finet's Ratio & -0.672 \\
Finet's   Ratio & -0.975 &               &        &               &       \\
\rowcolor{white}
\multicolumn{6}{p{18cm}}{\scriptsize{Note: Multiple linear regressions were performed with a backward method, whereby the subset with the lowest AIC was retained at each iteration. TAESS: Time Averaged Endothelial Shear Stress; OSI: Oscillatory Shear Index; and RRT: Relative Residence Time.}}
\end{tabular}
\end{table}

\section*{Discussion}
We have retrospectively analysed 127 CTCA scans of patients with suspected coronary artery disease and unremarkable stenosis or calcification. This constitutes the largest patient cohort, whereby all coronary bifurcation shape characteristics were correlated to haemodynamics.  In summary, our results show that a local haemodynamic can largely be explained with coronary shape characteristics:  (i) a smaller Finet's ratio (meaning the daughter branches are larger than common in comparison to the main vessel) significantly correlated to larger areas of adversely lowTAESS; (ii) larger branch diameters and stronger curvatures increased highOSI; and (iii) small Finet's ratio (daughter branches notably smaller than the main branch), large LAD and LCx diameter, and stronger curvatures each also increases highRRT. It should be noted that Finet's ratio showed the lowest variation within the cohort, suggesting small changes in Finet's ratio have a substantial impact on haemodynamics.

These metrics are strongly associated with the formation and progression of plaques as documented and evidenced by both \textit{in vitro} \cite{sho2004arterial} and cellular pathophysiology \cite{davies2009hemodynamic} studies.

Left main branch diameters and curvatures showed an interdependent, significant effect which appears sex-specific.  
It should be noted that tortuosity measured as tortuosity index has often been reported for this purpose in clinical literature. However, this needs to be interpreted with caution because it was demonstrated that this can lead to inconsistent results due to a variation in heart sphericity introducing inaccuracies due to three-dimensionality not being captured, with the mean average branch curvature as a superior measure \cite{kashyap2022accuracy}. 

We found that lower curvature can be considered favourable. However, higher incidence of both highOSI and highRRT still occurred in patients who also had larger branch diameters. Both highOSI and highRRT indicates stronger oscillatory shear stress (meaning topically periodically recurring), likely due to larger vessel diameter. This would suggest that a larger vessel diameter has a more impactful adverse effect than curvature, or perhaps that the adverse haemodynamic effect of larger diameters may be mitigated by lower curvature. Similarly, larger curvature appears to induce less highOSI and highRRT with fewer oscillatory blood flow dynamics if present with smaller branch diameters (both prominent in females). This indicates a potentially atheroprotective dynamic between coronary vessel diameter and curvature. Sex-specific haemodynamics are further explored elsewhere \cite{gharleghi2023sex}. This has not been described in the literature before, however, recent work hypothesised the interdependent significance of diameter with curvature in idealised work \cite{shen2022helical}.

No significant differences between the smoking versus non-smoking and hypertensive versus normal non-hypertensive subgroups were found after adjusting for multiple comparison testing. This potentially indicates that smoking and hypertension may not affect arterial remodelling in patients with coronary artery disease symptoms but without notable stenosis or calcification. Prior literature on this is inconsistent, however, with some work reporting smoking and hypertension correlated to increased tortuosity \cite{li2011clinical}, while others found no effect \cite{groves2009severe}. It is possible that the differences are too small to be detected at the statistical power in this study and future studies are warranted.

The cohort included only symptomatic patients in the absence of significant stenosis and calcium. Whether our findings would apply to a population with stenosis onset could not be stratified from our results and warrants future studies. A wider demographic representation may also provide additional valuable insights since increasing reports in literature point towards important population cohort differences, including ethnicities \cite{skowronski2020inter}, and age \cite{deopujari2010study}. 
This study relies on commonly used threshold-based haemodynamic assessment criteria. Still, pathophysiological processes ruling coronary artery disease have been found to be a continuum rather than a specific cut-off \cite{ku1997blood} and may be considered further in future studies. However, we have demonstrated that specific thresholds of the metrics considered did not affect the result. Mechanical forces within the vasculature are not considered here but may have an important effect  \cite{carpenter2020review}.  
The specific translation of this work into clinical practice must be further determined via clinical trials to validate the predictive power within the computational assumptions and simplifications made here.

Overall, our work sheds light on using coronary anatomical features to link and predict adverse local haemodynamics in the coronary left main bifurcation without stenosis. The results are promising in that some coronary anatomy characteristics are found to directly affect key pathophysiological drivers. We show that well-established indicators for adverse endothelial pathophysiological processes (TAESS, OSI, and RRT) can be largely predicted from geometrical features of the left main bifurcation, thereby allowing calculations of these risk metrics without computationally expensive CFD simulations. This may allow better patient-specific considerations of risk of coronary disease development in the future. 

\bibliography{refs}

\begin{thebibliography}{10}
\urlstyle{rm}
\expandafter\ifx\csname url\endcsname\relax
  \def\url#1{\texttt{#1}}\fi
\expandafter\ifx\csname urlprefix\endcsname\relax\def\urlprefix{URL }\fi
\expandafter\ifx\csname doiprefix\endcsname\relax\def\doiprefix{DOI: }\fi
\providecommand{\bibinfo}[2]{#2}
\providecommand{\eprint}[2][]{\url{#2}}

\bibitem{malek1999hemodynamic}
\bibinfo{author}{Malek, A.}, \bibinfo{author}{Alper, S.} \&
  \bibinfo{author}{Izumo, S.}
\newblock \bibinfo{journal}{\bibinfo{title}{Hemodynamic shear stress and its
  role in atherosclerosis}}.
\newblock {\emph{\JournalTitle{JAMA: The Journal of the American Medical
  Association}}} \textbf{\bibinfo{volume}{282}}, \bibinfo{pages}{2035}
  (\bibinfo{year}{1999}).

\bibitem{davies1995flow}
\bibinfo{author}{Davies, P.~F.}
\newblock \bibinfo{journal}{\bibinfo{title}{Flow-mediated endothelial
  mechanotransduction}}.
\newblock {\emph{\JournalTitle{Physiological reviews}}}
  \textbf{\bibinfo{volume}{75}}, \bibinfo{pages}{519--560}
  (\bibinfo{year}{1995}).

\bibitem{cecchi2011role}
\bibinfo{author}{Cecchi, E.} \emph{et~al.}
\newblock \bibinfo{journal}{\bibinfo{title}{Role of hemodynamic shear stress in
  cardiovascular disease}}.
\newblock {\emph{\JournalTitle{Atherosclerosis}}}
  \textbf{\bibinfo{volume}{214}}, \bibinfo{pages}{249--256},
  \doiprefix\url{10.1016/j.atherosclerosis.2010.09.008} (\bibinfo{year}{2011}).

\bibitem{loop1979atherosclerosis}
\bibinfo{author}{Loop, F.~D.} \emph{et~al.}
\newblock \bibinfo{journal}{\bibinfo{title}{Atherosclerosis of the left main
  coronary artery: 5 year results of surgical treatment}}.
\newblock {\emph{\JournalTitle{The American Journal of Cardiology}}}
  \textbf{\bibinfo{volume}{44}}, \bibinfo{pages}{195--201}
  (\bibinfo{year}{1979}).

\bibitem{zhu2009cataloguing}
\bibinfo{author}{Zhu, H.}, \bibinfo{author}{Ding, Z.}, \bibinfo{author}{Piana,
  R.~N.}, \bibinfo{author}{Gehrig, T.~R.} \& \bibinfo{author}{Friedman, M.~H.}
\newblock \bibinfo{journal}{\bibinfo{title}{Cataloguing the geometry of the
  human coronary arteries: a potential tool for predicting risk of coronary
  artery disease}}.
\newblock {\emph{\JournalTitle{International journal of cardiology}}}
  \textbf{\bibinfo{volume}{135}}, \bibinfo{pages}{43--52}
  (\bibinfo{year}{2009}).

\bibitem{morbiducci2016atherosclerosis}
\bibinfo{author}{Morbiducci, U.} \emph{et~al.}
\newblock \bibinfo{journal}{\bibinfo{title}{Atherosclerosis at arterial
  bifurcations: evidence for the role of haemodynamics and geometry}}.
\newblock {\emph{\JournalTitle{Journal of Thrombosis and Haemostasis}}}
  \textbf{\bibinfo{volume}{115}}, \bibinfo{pages}{484--92},
  \doiprefix\url{10.1160/TH15-07-0597} (\bibinfo{year}{2016}).

\bibitem{beier2016dynamically}
\bibinfo{author}{Beier, S.} \emph{et~al.}
\newblock \bibinfo{journal}{\bibinfo{title}{Dynamically scaled phantom phase
  contrast mri compared to true‐scale computational modeling of coronary
  artery flow}}.
\newblock {\emph{\JournalTitle{Journal of Magnetic Resonance Imaging}}}
  (\bibinfo{year}{2016}).

\bibitem{taylor2004experimental}
\bibinfo{author}{Taylor, C.~A.} \& \bibinfo{author}{Draney, M.~T.}
\newblock \emph{\bibinfo{title}{Experimental and computational methods in
  cardiovascular fluid mechanics}}, vol.~\bibinfo{volume}{36}
  (\bibinfo{publisher}{Annual Reviews}, \bibinfo{year}{2004}).

\bibitem{sitzer2003internal}
\bibinfo{author}{Sitzer, M.} \emph{et~al.}
\newblock \bibinfo{journal}{\bibinfo{title}{Internal carotid artery angle of
  origin: a novel risk factor for early carotid atherosclerosis}}.
\newblock {\emph{\JournalTitle{Stroke}}} \textbf{\bibinfo{volume}{34}},
  \bibinfo{pages}{950--955} (\bibinfo{year}{2003}).

\bibitem{zhang2012flow}
\bibinfo{author}{Zhang, C.} \emph{et~al.}
\newblock \bibinfo{journal}{\bibinfo{title}{Flow patterns and wall shear stress
  distribution in human internal carotid arteries: the geometric effect on the
  risk for stenoses}}.
\newblock {\emph{\JournalTitle{Journal of biomechanics}}}
  \textbf{\bibinfo{volume}{45}}, \bibinfo{pages}{83--89}
  (\bibinfo{year}{2012}).

\bibitem{lee2008geometry}
\bibinfo{author}{Lee, S.-W.}, \bibinfo{author}{Antiga, L.},
  \bibinfo{author}{Spence, J.~D.} \& \bibinfo{author}{Steinman, D.~A.}
\newblock \bibinfo{journal}{\bibinfo{title}{Geometry of the carotid bifurcation
  predicts its exposure to disturbed flow}}.
\newblock {\emph{\JournalTitle{Stroke}}} \textbf{\bibinfo{volume}{39}},
  \bibinfo{pages}{2341--2347} (\bibinfo{year}{2008}).

\bibitem{fedintsev2017markers}
\bibinfo{author}{Fedintsev, A.} \emph{et~al.}
\newblock \bibinfo{journal}{\bibinfo{title}{Markers of arterial health could
  serve as accurate non-invasive predictors of human biological and
  chronological age}}.
\newblock {\emph{\JournalTitle{Aging (Albany NY)}}}
  \textbf{\bibinfo{volume}{9}}, \bibinfo{pages}{1280} (\bibinfo{year}{2017}).

\bibitem{o1996thickening}
\bibinfo{author}{O’Leary, D.~H.} \emph{et~al.}
\newblock \bibinfo{journal}{\bibinfo{title}{Thickening of the carotid wall: a
  marker for atherosclerosis in the elderly?}}
\newblock {\emph{\JournalTitle{Stroke}}} \textbf{\bibinfo{volume}{27}},
  \bibinfo{pages}{224--231} (\bibinfo{year}{1996}).

\bibitem{liebeskind2015carotid}
\bibinfo{author}{Liebeskind, D.~S.} \emph{et~al.}
\newblock \bibinfo{journal}{\bibinfo{title}{Carotid i's, l's and t's:
  collaterals shape the outcome of intracranial carotid occlusion in acute
  ischemic stroke}}.
\newblock {\emph{\JournalTitle{Journal of neurointerventional surgery}}}
  \textbf{\bibinfo{volume}{7}}, \bibinfo{pages}{402--407}
  (\bibinfo{year}{2015}).

\bibitem{beier2016impact}
\bibinfo{author}{Beier, S.} \emph{et~al.}
\newblock \bibinfo{journal}{\bibinfo{title}{Impact of bifurcation angle and
  other anatomical characteristics on blood flow–a computational study of
  non-stented and stented coronary arteries}}.
\newblock {\emph{\JournalTitle{Journal of Biomechanics}}}
  \textbf{\bibinfo{volume}{49}}, \bibinfo{pages}{1570--1582}
  (\bibinfo{year}{2016}).

\bibitem{shen2021secondary}
\bibinfo{author}{Shen, C.} \emph{et~al.}
\newblock \bibinfo{journal}{\bibinfo{title}{Secondary flow in bifurcations –
  important effects of curvature, bifurcation angle and stents}}.
\newblock {\emph{\JournalTitle{Journal of Biomechanics}}}
  \textbf{\bibinfo{volume}{129}}, \bibinfo{pages}{110755},
  \doiprefix\url{https://doi.org/10.1016/j.jbiomech.2021.110755}
  (\bibinfo{year}{2021}).

\bibitem{rabbi2020computational}
\bibinfo{author}{Rabbi, M.~F.}, \bibinfo{author}{Laboni, F.~S.} \&
  \bibinfo{author}{Arafat, M.~T.}
\newblock \bibinfo{journal}{\bibinfo{title}{Computational analysis of the
  coronary artery hemodynamics with different anatomical variations}}.
\newblock {\emph{\JournalTitle{Informatics in Medicine Unlocked}}}
  \textbf{\bibinfo{volume}{19}}, \doiprefix\url{10.1016/j.imu.2020.100314}
  (\bibinfo{year}{2020}).

\bibitem{li2011clinical}
\bibinfo{author}{Li, Y.} \emph{et~al.}
\newblock \bibinfo{journal}{\bibinfo{title}{Clinical implication of coronary
  tortuosity in patients with coronary artery disease}}.
\newblock {\emph{\JournalTitle{PloS one}}} \textbf{\bibinfo{volume}{6}},
  \bibinfo{pages}{e24232} (\bibinfo{year}{2011}).

\bibitem{tuncay2018non}
\bibinfo{author}{Tuncay, V.} \emph{et~al.}
\newblock \bibinfo{journal}{\bibinfo{title}{Non-invasive assessment of coronary
  artery geometry using coronary cta}}.
\newblock {\emph{\JournalTitle{Journal of Cardiovascular Computed Tomography}}}
  \textbf{\bibinfo{volume}{12}}, \bibinfo{pages}{257--260},
  \doiprefix\url{10.1016/J.JCCT.2018.02.003} (\bibinfo{year}{2018}).

\bibitem{li2017impact}
\bibinfo{author}{Li, Y.} \emph{et~al.}
\newblock \bibinfo{journal}{\bibinfo{title}{Impact of coronary tortuosity on
  coronary pressure and wall shear stress: an experimental study}}.
\newblock {\emph{\JournalTitle{Mol Cell Biomech}}}
  \textbf{\bibinfo{volume}{14}}, \bibinfo{pages}{213--9}
  (\bibinfo{year}{2017}).

\bibitem{kashyap2022accuracy}
\bibinfo{author}{Kashyap, V.} \emph{et~al.}
\newblock \bibinfo{journal}{\bibinfo{title}{Accuracy of vascular tortuosity
  measures using computational modelling}}.
\newblock {\emph{\JournalTitle{Scientific reports}}}
  \textbf{\bibinfo{volume}{12}}, \bibinfo{pages}{1--10} (\bibinfo{year}{2022}).

\bibitem{xie2014computation}
\bibinfo{author}{Xie, X.}, \bibinfo{author}{Wang, Y.}, \bibinfo{author}{Zhu,
  H.} \& \bibinfo{author}{Zhou, J.}
\newblock \bibinfo{journal}{\bibinfo{title}{Computation of hemodynamics in
  tortuous left coronary artery: A morphological parametric study}}.
\newblock {\emph{\JournalTitle{Journal of Biomechanical Engineering}}}
  \textbf{\bibinfo{volume}{136}}, \doiprefix\url{10.1115/1.4028052/370058}
  (\bibinfo{year}{2014}).

\bibitem{song2021numerical}
\bibinfo{author}{Song, J.}, \bibinfo{author}{Kouidri, S.} \&
  \bibinfo{author}{Bakir, F.}
\newblock \bibinfo{journal}{\bibinfo{title}{Numerical study on flow topology
  and hemodynamics in tortuous coronary artery with symmetrical and
  asymmetrical stenosis}}.
\newblock {\emph{\JournalTitle{Biocybernetics and Biomedical Engineering}}}
  \textbf{\bibinfo{volume}{41}}, \bibinfo{pages}{142--155},
  \doiprefix\url{10.1016/J.BBE.2020.12.006} (\bibinfo{year}{2021}).

\bibitem{muller2012pressure}
\bibinfo{author}{Muller, O.} \emph{et~al.}
\newblock \bibinfo{journal}{\bibinfo{title}{Pressure–diameter relationship in
  human coronary arteries}}.
\newblock {\emph{\JournalTitle{Circulation: Cardiovascular Interventions}}}
  \textbf{\bibinfo{volume}{5}}, \bibinfo{pages}{791--796},
  \doiprefix\url{10.1161/CIRCINTERVENTIONS.112.972224} (\bibinfo{year}{2012}).

\bibitem{shen2022helical}
\bibinfo{author}{Shen, C.}, \bibinfo{author}{Gharleghi, R.},
  \bibinfo{author}{Li, D.} \& \bibinfo{author}{Beier, S.}
\newblock \bibinfo{title}{Helical flow in healthy and diseased patient-specific
  coronary bifurcations}.
\newblock In \emph{\bibinfo{booktitle}{2022 44th Annual International
  Conference of the IEEE Engineering in Medicine \& Biology Society (EMBC)}},
  \bibinfo{pages}{3977--3980} (\bibinfo{organization}{IEEE},
  \bibinfo{year}{2022}).

\bibitem{pinho2019impact}
\bibinfo{author}{Pinho, N.} \emph{et~al.}
\newblock \bibinfo{journal}{\bibinfo{title}{The impact of the right coronary
  artery geometric parameters on hemodynamic performance}}.
\newblock {\emph{\JournalTitle{Cardiovascular engineering and technology}}}
  \textbf{\bibinfo{volume}{10}}, \bibinfo{pages}{257--270}
  (\bibinfo{year}{2019}).

\bibitem{medrano_gracia2016study}
\bibinfo{author}{Medrano-Gracia, P.} \emph{et~al.}
\newblock \bibinfo{journal}{\bibinfo{title}{A study of coronary bifurcation
  shape in a normal population}}.
\newblock {\emph{\JournalTitle{Journal of Cardiovascular Translational
  Research}}} \bibinfo{pages}{1--9} (\bibinfo{year}{2016}).

\bibitem{medrano_gracia2014construction}
\bibinfo{author}{Medrano-Gracia, P.} \emph{et~al.}
\newblock \bibinfo{journal}{\bibinfo{title}{Construction of a coronary atlas
  from ct angiography}}.
\newblock {\emph{\JournalTitle{MICCAI}}}  (\bibinfo{year}{2014}).

\bibitem{piccinelli2009framework}
\bibinfo{author}{Piccinelli, M.}, \bibinfo{author}{Veneziani, A.},
  \bibinfo{author}{Steinman, D.~A.}, \bibinfo{author}{Remuzzi, A.} \&
  \bibinfo{author}{Antiga, L.}
\newblock \bibinfo{journal}{\bibinfo{title}{A framework for geometric analysis
  of vascular structures: application to cerebral aneurysms}}.
\newblock {\emph{\JournalTitle{IEEE transactions on medical imaging}}}
  \textbf{\bibinfo{volume}{28}}, \bibinfo{pages}{1141--1155}
  (\bibinfo{year}{2009}).

\bibitem{taubincurve}
\bibinfo{author}{Taubin, G.}
\newblock \bibinfo{title}{Curve and surface smoothing without shrinkage}.
\newblock In \emph{\bibinfo{booktitle}{Proceedings of IEEE international
  conference on computer vision}}, \bibinfo{pages}{852--857}
  (\bibinfo{organization}{IEEE}, \bibinfo{year}{1995}).

\bibitem{lassen2014percutaneous}
\bibinfo{author}{Lassen, J.~F.} \emph{et~al.}
\newblock \bibinfo{journal}{\bibinfo{title}{Percutaneous coronary intervention
  for coronary bifurcation disease: consensus from the first 10 years of the
  european bifurcation club meetings}}.
\newblock {\emph{\JournalTitle{EuroIntervention}}}
  \textbf{\bibinfo{volume}{10}}, \bibinfo{pages}{545--560}
  (\bibinfo{year}{2014}).

\bibitem{eslami2020effect}
\bibinfo{author}{Eslami, P.} \emph{et~al.}
\newblock \bibinfo{journal}{\bibinfo{title}{Effect of wall elasticity on
  hemodynamics and wall shear stress in patient-specific simulations in the
  coronary arteries}}.
\newblock {\emph{\JournalTitle{Journal of biomechanical engineering}}}
  \textbf{\bibinfo{volume}{142}}, \bibinfo{pages}{024503}
  (\bibinfo{year}{2020}).

\bibitem{tangasme}
\bibinfo{author}{Tang, T.}, \bibinfo{author}{Giddens, D.},
  \bibinfo{author}{Zarins, C.} \& \bibinfo{author}{Glagov, S.}
\newblock \bibinfo{title}{Velocity profile and wall shear measurements in a
  model human coronary artery}.
\newblock In \emph{\bibinfo{booktitle}{ASME Winter Annual Meeting Proceedings,
  Atlanta. New York: ASME}} (\bibinfo{year}{1991}).

\bibitem{razavi2011numerical}
\bibinfo{author}{Razavi, A.}, \bibinfo{author}{Shirani, E.} \&
  \bibinfo{author}{Sadeghi, M.~R.}
\newblock \bibinfo{journal}{\bibinfo{title}{Numerical simulation of blood
  pulsatile flow in a stenosed carotid artery using different rheological
  models}}.
\newblock {\emph{\JournalTitle{Journal of Biomechanics}}}
  \textbf{\bibinfo{volume}{44}}, \bibinfo{pages}{2021--30}
  (\bibinfo{year}{2011}).

\bibitem{he1996pulsatile}
\bibinfo{author}{He, X.} \& \bibinfo{author}{Ku, D.~N.}
\newblock \bibinfo{journal}{\bibinfo{title}{Pulsatile flow in the human left
  coronary artery bifurcation: average conditions}}.
\newblock {\emph{\JournalTitle{Journal of biomechanical engineering}}}
  \textbf{\bibinfo{volume}{118}}, \bibinfo{pages}{74--82}
  (\bibinfo{year}{1996}).

\bibitem{hoi2011correlation}
\bibinfo{author}{Hoi, Y.}, \bibinfo{author}{Zhou, Y.-Q.},
  \bibinfo{author}{Zhang, X.}, \bibinfo{author}{Henkelman, R.~M.} \&
  \bibinfo{author}{Steinman, D.~A.}
\newblock \bibinfo{journal}{\bibinfo{title}{Correlation between local
  hemodynamics and lesion distribution in a novel aortic regurgitation murine
  model of atherosclerosis}}.
\newblock {\emph{\JournalTitle{Annals of biomedical engineering}}}
  \textbf{\bibinfo{volume}{39}}, \bibinfo{pages}{1414--1422}
  (\bibinfo{year}{2011}).

\bibitem{soulis2014wall}
\bibinfo{author}{Soulis, J.}, \bibinfo{author}{Fytanidis, D.},
  \bibinfo{author}{Seralidou, K.} \& \bibinfo{author}{Giannoglou, G.}
\newblock \bibinfo{journal}{\bibinfo{title}{Wall shear stress oscillation and
  its gradient in the normal left coronary artery tree bifurcations}}.
\newblock {\emph{\JournalTitle{Hippokratia}}} \textbf{\bibinfo{volume}{18}},
  \bibinfo{pages}{12} (\bibinfo{year}{2014}).

\bibitem{seabold}
\bibinfo{author}{Seabold, S.} \& \bibinfo{author}{Perktold, J.}
\newblock \bibinfo{title}{Statsmodels: Econometric and statistical modeling
  with python}.
\newblock In \emph{\bibinfo{booktitle}{Proceedings of the 9th Python in Science
  Conference}}, vol.~\bibinfo{volume}{57}, \bibinfo{pages}{10--25080}
  (\bibinfo{organization}{Austin, TX}, \bibinfo{year}{2010}).

\bibitem{ruxton2006unequal}
\bibinfo{author}{Ruxton, G.~D.}
\newblock \bibinfo{journal}{\bibinfo{title}{The unequal variance t-test is an
  underused alternative to student's t-test and the mann–whitney u test}}.
\newblock {\emph{\JournalTitle{Behavioral Ecology}}}
  \textbf{\bibinfo{volume}{17}}, \bibinfo{pages}{688--690}
  (\bibinfo{year}{2006}).

\bibitem{delacre2017why}
\bibinfo{author}{Delacre, M.}, \bibinfo{author}{Lakens, D.} \&
  \bibinfo{author}{Leys, C.}
\newblock \bibinfo{journal}{\bibinfo{title}{Why psychologists should by default
  use welch’s t-test instead of student’s t-test}}.
\newblock {\emph{\JournalTitle{International Review of Social Psychology}}}
  \textbf{\bibinfo{volume}{30}} (\bibinfo{year}{2017}).

\bibitem{abdi2010holm}
\bibinfo{author}{Abdi, H.}
\newblock \bibinfo{journal}{\bibinfo{title}{Holm’s sequential bonferroni
  procedure}}.
\newblock {\emph{\JournalTitle{Encyclopedia of research design}}}
  \textbf{\bibinfo{volume}{1}}, \bibinfo{pages}{1--8} (\bibinfo{year}{2010}).

\bibitem{yamashita2007stepwise}
\bibinfo{author}{Yamashita, T.}, \bibinfo{author}{Yamashita, K.} \&
  \bibinfo{author}{Kamimura, R.}
\newblock \bibinfo{journal}{\bibinfo{title}{A stepwise aic method for variable
  selection in linear regression}}.
\newblock {\emph{\JournalTitle{Communications in Statistics-Theory and
  Methods}}}  (\bibinfo{year}{2007}).

\bibitem{hu2007akaike}
\bibinfo{author}{Hu, S.}
\newblock \bibinfo{journal}{\bibinfo{title}{Akaike information criterion}}.
\newblock {\emph{\JournalTitle{Center for Research in Scientific Computation}}}
  \textbf{\bibinfo{volume}{93}} (\bibinfo{year}{2007}).

\bibitem{sho2004arterial}
\bibinfo{author}{Sho, E.} \emph{et~al.}
\newblock \bibinfo{journal}{\bibinfo{title}{Arterial enlargement, tortuosity,
  and intimal thickening in response to sequential exposure to high and low
  wall shear stress}}.
\newblock {\emph{\JournalTitle{Journal of vascular surgery}}}
  \textbf{\bibinfo{volume}{39}}, \bibinfo{pages}{601--612}
  (\bibinfo{year}{2004}).

\bibitem{davies2009hemodynamic}
\bibinfo{author}{Davies, P.~F.}
\newblock \bibinfo{journal}{\bibinfo{title}{Hemodynamic shear stress and the
  endothelium in cardiovascular pathophysiology}}.
\newblock {\emph{\JournalTitle{Nature clinical practice Cardiovascular
  medicine}}} \textbf{\bibinfo{volume}{6}}, \bibinfo{pages}{16--26}
  (\bibinfo{year}{2009}).

\bibitem{gharleghi2023sex}
\bibinfo{author}{Gharleghi, R.} \emph{et~al.}
\newblock \bibinfo{journal}{\bibinfo{title}{Sex-specific variances in anatomy
  and blood flow of the left main coronary bifurcation: Implications for
  coronary artery disease risk}}.
\newblock {\emph{\JournalTitle{arXiv preprint arXiv:2311.18489}}}
  (\bibinfo{year}{2023}).

\bibitem{groves2009severe}
\bibinfo{author}{Groves, S.~S.} \emph{et~al.}
\newblock \bibinfo{journal}{\bibinfo{title}{Severe coronary tortuosity and the
  relationship to significant coronary artery disease}}.
\newblock {\emph{\JournalTitle{West Virginia Medical Journal}}}
  \textbf{\bibinfo{volume}{105}}, \bibinfo{pages}{14--18}
  (\bibinfo{year}{2009}).

\bibitem{skowronski2020inter}
\bibinfo{author}{Skowronski, J.} \emph{et~al.}
\newblock \bibinfo{journal}{\bibinfo{title}{Inter-ethnic differences in normal
  coronary anatomy between caucasian (polish) and asian (korean) populations}}.
\newblock {\emph{\JournalTitle{European Journal of Radiology}}}
  \textbf{\bibinfo{volume}{130}}, \bibinfo{pages}{109185}
  (\bibinfo{year}{2020}).

\bibitem{deopujari2010study}
\bibinfo{author}{Deopujari, R.} \& \bibinfo{author}{Dixit, A.}
\newblock \bibinfo{journal}{\bibinfo{title}{The study of age related changes in
  coronary arteries and its relevance to the atherosclerosis}}.
\newblock {\emph{\JournalTitle{Journal of Anatomical Society of India}}}
  \textbf{\bibinfo{volume}{59}}, \bibinfo{pages}{192--196}
  (\bibinfo{year}{2010}).

\bibitem{ku1997blood}
\bibinfo{author}{Ku, D.}
\newblock \bibinfo{journal}{\bibinfo{title}{Blood flow in arteries}}.
\newblock {\emph{\JournalTitle{Annual Review of Fluid Mechanics}}}
  \textbf{\bibinfo{volume}{29}}, \bibinfo{pages}{399--434}
  (\bibinfo{year}{1997}).

\bibitem{carpenter2020review}
\bibinfo{author}{Carpenter, H.~J.}, \bibinfo{author}{Gholipour, A.},
  \bibinfo{author}{Ghayesh, M.~H.}, \bibinfo{author}{Zander, A.~C.} \&
  \bibinfo{author}{Psaltis, P.~J.}
\newblock \bibinfo{journal}{\bibinfo{title}{A review on the biomechanics of
  coronary arteries}}.
\newblock {\emph{\JournalTitle{International Journal of Engineering Science}}}
  \textbf{\bibinfo{volume}{147}}, \bibinfo{pages}{103201}
  (\bibinfo{year}{2020}).

\end{thebibliography}

\section*{Acknowledgements}
SB acknowledges Intra for assistance in collection of the dataset and the Auckland Academic Health Alliance (AAHA) and the Auckland Medical Research Foundation (AMRF) for their financial support and endorsement.

This research was undertaken with the assistance of computational resources from the National Computational Infrastructure (NCI), which is supported by the Australian Government, and the computational cluster Katana supported by Research Technology Services at UNSW Sydney.

\section*{Author contributions statement}
RG designed and performed the haemodynamic simulations and analysed the results with assistance from MZ, CS and SB. CE and MW assisted with the data collection and interpretation of results. All authors contributed to the preparation of this manuscript.

\section*{Additional information}
\paragraph{Competing interests} The authors declare that they have no conflicts of interest.

\paragraph{Data Availability} The datasets generated and analyzed during the current study are available from the corresponding author on reasonable request.
\newpage
\appendix
\section*{Supplementary Material}
\renewcommand{\thetable}{A.\arabic{table}}
\renewcommand{\thefigure}{A.\arabic{figure}}
\setcounter{figure}{0}    
\setcounter{table}{0}    
\begin{table}[h]
\centering
\caption{Influence of sex on anatomical and haemodynamic metrics studied.}
\label{tab:sexsupp}
\rowcolors{2}{gray!25}{white}
\begin{tabular}{@{}lcccccc@{}}
\textbf{Feature} & \textbf{t-statistic} & \textbf{p-value} & \textbf{Difference} & \multicolumn{2}{l}{\textbf{95\% CI}} & \textbf{Adjusted p-value} \\\midrule
Angle A       & 0.699  & 0.486            & 2.465  & -4.536  & 9.467  & 1                \\
Angle B       & 1.117  & 0.267            & 3.959  & -3.096  & 11.014 & 1                \\
Angle C       & -1.172 & 0.244            & -3.392 & -9.122  & 2.338  & 1                \\
Inflow Angle  & -2.266 & 0.025            & -5.571 & -10.444 & -0.698 & 0.175            \\
LM Curvature  & -3.341 & 0.001            & -0.062 & -0.099  & -0.025 & 0.008            \\
LAD Curvature & -3.872 & \textless{}0.001 & -0.063 & -0.096  & -0.031 & \textless{}0.001* \\
LCX Curvature & -3.738 & \textless{}0.001 & -0.072 & -0.11   & -0.034 & \textless{}0.001* \\
LM Diameter   & 6.614  & \textless{}0.001 & 0.691  & 0.483   & 0.898  & \textless{}0.001* \\
LAD Diameter  & 6.352  & \textless{}0.001 & 0.605  & 0.415   & 0.794  & \textless{}0.001* \\
LCX Diameter  & 6.081  & \textless{}0.001 & 0.631  & 0.425   & 0.837  & \textless{}0.001* \\
Finet's ratio & 0.021  & 0.984            & 0      & -0.01   & 0.011  & 1                \\
Low TAESS     & 1.101  & 0.274            & 1.818  & -1.459  & 5.096  & 0.274            \\
High OSI      & 4.688  & \textless{}0.001 & 2.225  & 1.274   & 3.175  & \textless{}0.001* \\
High RRT      & 2.927  & 0.005            & 1.823  & 0.578   & 3.067  & 0.01            
\end{tabular}
\end{table}

\begin{table}[h]
\centering
\caption{Influence of hypertension on anatomical and haemodynamic metrics studied.}
\label{tab:hypersupp}
\rowcolors{2}{gray!25}{white}
\begin{tabular}{@{}lcccccc@{}}
\textbf{Feature} & \textbf{t-statistic} & \textbf{p-value} & \textbf{Difference} & \multicolumn{2}{l}{\textbf{95\% CI}} & \textbf{Adjusted p-value} \\\midrule
Angle A       & 0.795  & 0.43  & 3.207  & -4.858  & 11.272 & 1     \\
Angle B       & -2.149 & 0.035 & -7.692 & -14.836 & -0.548 & 0.49  \\
Angle C       & 2.049  & 0.044 & 7.045  & 0.194   & 13.896 & 0.528 \\
Inflow Angle  & 0.036  & 0.972 & 0.114  & -6.296  & 6.524  & 1     \\
LM Curvature  & 0.491  & 0.625 & 0.011  & -0.032  & 0.053  & 1     \\
LAD Curvature & 1.074  & 0.287 & 0.021  & -0.018  & 0.06   & 1     \\
LCX Curvature & 1.444  & 0.153 & 0.034  & -0.013  & 0.08   & 1     \\
LM Diameter   & -0.951 & 0.345 & -0.122 & -0.379  & 0.134  & 1     \\
LAD Diameter  & -1.152 & 0.253 & -0.127 & -0.346  & 0.093  & 1     \\
LCX Diameter  & -1.455 & 0.151 & -0.187 & -0.445  & 0.07   & 1     \\
Finet's ratio & 1.164  & 0.248 & 0.007  & -0.005  & 0.019  & 1     \\
Low TAESS     & -0.531 & 0.596 & -0.848 & -4.017  & 2.321  & 1     \\
High OSI      & -0.623 & 0.535 & -0.266 & -1.115  & 0.584  & 1     \\
High RRT      & -0.505 & 0.615 & -0.273 & -1.35   & 0.803  & 1    
\end{tabular}
\end{table}

\begin{table}[h]
\centering
\caption{Influence of smoking history on anatomical and haemodynamic metrics studied.}
\label{tab:smokingsupp}
\rowcolors{2}{gray!25}{white}
\begin{tabular}{@{}lcccccc@{}}
\textbf{Feature} & \textbf{t-statistic} & \textbf{p-value} & \textbf{Difference} & \multicolumn{2}{l}{\textbf{95\% CI}} & \textbf{Adjusted p-value} \\\midrule
Angle A       & -1.024 & 0.309 & -4.127 & -12.171 & 3.916  & 1     \\
Angle B       & 1.048  & 0.298 & 3.897  & -3.516  & 11.31  & 1     \\
Angle C       & 0.212  & 0.833 & 0.693  & -5.805  & 7.191  & 1     \\
Inflow Angle  & 0.871  & 0.386 & 2.412  & -3.099  & 7.923  & 1     \\
LM Curvature  & 0.416  & 0.679 & 0.008  & -0.032  & 0.049  & 1     \\
LAD Curvature & 0.238  & 0.812 & 0.004  & -0.032  & 0.041  & 1     \\
LCX Curvature & -0.16  & 0.874 & -0.004 & -0.048  & 0.041  & 1     \\
LM Diameter   & -0.678 & 0.499 & -0.077 & -0.304  & 0.149  & 1     \\
LAD Diameter  & -0.095 & 0.925 & -0.01  & -0.211  & 0.192  & 1     \\
LCX Diameter  & 0.182  & 0.856 & 0.02   & -0.202  & 0.243  & 1     \\
Finet's ratio & -2.527 & 0.013 & -0.015 & -0.027  & -0.003 & 0.182 \\
Low TAESS     & 1.535  & 0.129 & 2.697  & -0.799  & 6.192  & 0.218 \\
High OSI      & 1.615  & 0.109 & 0.648  & -0.148  & 1.443  & 0.218 \\
High RRT      & 2.228  & 0.028 & 1.165  & 0.128   & 2.203  & 0.084
\end{tabular}
\end{table}

\begin{landscape}
\begin{table}[h]
\centering
\caption{Impact of TAESS threshold of 0.4, 0.5 and 0.6Pa (left to right) on statistically significant variables with statistically significant findings* with p<0.05 marked in yellow, and green after Bonferroni correction.}
\label{tab:TAESSThreshold}
\rowcolors{2}{gray!25}{white}
\begin{tabular}{@{}l|cccC{1.5cm}|cccC{1.5cm}|cccC{1.5cm}@{}}
 & \multicolumn{4}{c}{\textbf{LowTAESS \textless 0.4 Pa}} & \multicolumn{4}{c}{\textbf{LowTAESS \textless 0.5 Pa}} & \multicolumn{4}{c}{\textbf{LowTAESS\textless{}0.6 Pa}} \\
\textbf{Feature} & R$^2$ & Coefficient & p-value & Adjusted p-value & R$^2$ & Coefficient & p-value & Adjusted p-value & R$^2$ & Coefficient & p-value & Adjusted p-value \\\midrule
Weight & 0.012 & -0.107 & 0.229 & 1 & 0.012 & -0.111 & 0.215 & 1 & 0.01 & -0.101 & 0.26 & 1 \\
BMI & 0.025 & -0.158 & 0.077 & 0.767 & 0.022 & -0.147 & 0.099 & 0.89 & 0.014 & -0.116 & 0.192 & 1 \\
Angle A & 0.001 & 0.035 & 0.7 & 1 & 0.001 & -0.028 & 0.753 & 1 & 0.002 & -0.049 & 0.581 & 1 \\
Angle B & 0.008 & -0.09 & 0.312 & 1 & 0.006 & -0.078 & 0.381 & 1 & 0.011 & -0.104 & 0.244 & 1 \\
Angle C & 0.008 & 0.088 & 0.324 & 1 & 0.026 & 0.161 & 0.071 & 0.78 & 0.048 & 0.219 & \cellcolor[HTML]{FFF2CC}0.014 & 0.163 \\
Inflow   Angle & 0 & -0.013 & 0.885 & 1 & 0.001 & 0.034 & 0.702 & 1 & 0.008 & 0.089 & 0.319 & 1 \\
LM   Curvature & 0.014 & 0.119 & 0.184 & 1 & 0.009 & 0.094 & 0.294 & 1 & 0.003 & 0.056 & 0.533 & 1 \\
LAD   Curvature & 0 & 0.006 & 0.947 & 1 & 0 & 0.006 & 0.948 & 1 & 0 & 0.002 & 0.978 & 1 \\
LCX   Curvature & 0.011 & -0.104 & 0.243 & 1 & 0.013 & -0.112 & 0.208 & 1 & 0.014 & -0.119 & 0.184 & 1 \\
LM   Diameter & 0 & 0.019 & 0.83 & 1 & 0.004 & -0.061 & 0.494 & 1 & 0.014 & -0.12 & 0.179 & 1 \\
LAD   Diameter & 0.054 & 0.232 & \cellcolor[HTML]{FFF2CC}0.009 & 0.106 & 0.026 & 0.16 & 0.072 & 0.78 & 0.008 & 0.091 & 0.311 & 1 \\
LCX   Diameter & 0.048 & 0.219 & \cellcolor[HTML]{FFF2CC}0.013 & 0.148 & 0.03 & 0.172 & 0.053 & 0.634 & 0.016 & 0.125 & 0.162 & 1 \\
Finet & 0.598 & -0.773 & \cellcolor[HTML]{FFF2CC}\textless{}0.001 & \cellcolor[HTML]{C6E0B4}\textless{}0.001* & 0.709 & -0.842 & \cellcolor[HTML]{FFF2CC}\textless{}0.001 & \cellcolor[HTML]{C6E0B4}\textless{}0.001* & 0.704 & -0.839 & \cellcolor[HTML]{FFF2CC}\textless{}0.001 & \cellcolor[HTML]{C6E0B4}\textless{}0.001*
\end{tabular}
\end{table}
\end{landscape}

\begin{table}[h]
\centering
\caption{Impact of OSI threshold on statistically significant variables with * signifying p<0.05 marked in yellow, and green after Bonreffoni correction. }
\label{tab:OSIThresholds}
\rowcolors{2}{gray!25}{white}
\begin{tabular}{@{}l|cccC{1.5cm}|cccC{1.5cm}@{}}
 & \multicolumn{4}{c}{\textbf{HighOSI\textgreater{}0.1}} & \multicolumn{4}{c}{\textbf{HighOSI\textgreater{}0.2}} \\
\textbf{Feature} & R$^2$ & Coefficient & p-value & Adjusted p-value & R$^2$ & Coefficient & p-value & Adjusted p-value \\\midrule
Weight & 0.027 & 0.164 & 0.066 & 0.397 & 0.015 & 0.121 & 0.174 & 1 \\
BMI & 0.007 & -0.082 & 0.362 & 1 & 0.011 & -0.104 & 0.242 & 1 \\
Angle A & 0 & 0.02 & 0.826 & 1 & 0 & 0.013 & 0.885 & 1 \\
Angle B & 0.001 & 0.027 & 0.766 & 1 & 0 & -0.015 & 0.865 & 1 \\
Angle C & 0.002 & -0.048 & 0.593 & 1 & 0 & 0.002 & 0.98 & 1 \\
Inflow   Angle & 0.001 & -0.038 & 0.673 & 1 & 0 & 0.006 & 0.944 & 1 \\
LM   Curvature & 0.134 & -0.367 & \cellcolor[HTML]{FFF2CC}\textless{}0.001 & \cellcolor[HTML]{C6E0B4}\textless{}0.001* & 0.092 & -0.304 & \cellcolor[HTML]{FFF2CC}0.001 & \cellcolor[HTML]{C6E0B4}0.004* \\
LAD   Curvature & 0.234 & -0.484 & \cellcolor[HTML]{FFF2CC}\textless{}0.001 & \cellcolor[HTML]{C6E0B4}\textless{}0.001* & 0.171 & -0.413 & \cellcolor[HTML]{FFF2CC}\textless{}0.001 & \cellcolor[HTML]{C6E0B4}\textless{}0.001* \\
LCX   Curvature & 0.27 & -0.519 & \cellcolor[HTML]{FFF2CC}\textless{}0.001 & \cellcolor[HTML]{C6E0B4}\textless{}0.001* & 0.158 & -0.397 & \cellcolor[HTML]{FFF2CC}\textless{}0.001 & \cellcolor[HTML]{C6E0B4}\textless{}0.001* \\
LM   Diameter & 0.477 & 0.691 & \cellcolor[HTML]{FFF2CC}\textless{}0.001 & \cellcolor[HTML]{C6E0B4}\textless{}0.001* & 0.332 & 0.576 & \cellcolor[HTML]{FFF2CC}\textless{}0.001 & \cellcolor[HTML]{C6E0B4}\textless{}0.001* \\
LAD   Diameter & 0.545 & 0.738 & \cellcolor[HTML]{FFF2CC}\textless{}0.001 & \cellcolor[HTML]{C6E0B4}\textless{}0.001* & 0.416 & 0.645 & \cellcolor[HTML]{FFF2CC}\textless{}0.001 & \cellcolor[HTML]{C6E0B4}\textless{}0.001 \\
LCX   Diameter & 0.546 & 0.739 & \cellcolor[HTML]{FFF2CC}\textless{}0.001 & \cellcolor[HTML]{C6E0B4}\textless{}0.001* & 0.357 & 0.598 & \cellcolor[HTML]{FFF2CC}\textless{}0.001 & \cellcolor[HTML]{C6E0B4}\textless{}0.001* \\
Finet & 0.035 & -0.187 & \cellcolor[HTML]{FFF2CC}0.035 & 0.247 & 0.031 & -0.176 & \cellcolor[HTML]{FFF2CC}0.048 & 0.336
\end{tabular}
\end{table}
\end{document}